\documentclass[11pt]{article}
\usepackage{amsmath,amsfonts,amsthm} 
\usepackage{amssymb}

\newtheorem{theorem}{Theorem}
\newtheorem{lemma}[theorem]{Lemma}
\newtheorem{proposition}[theorem]{Proposition}

\newcommand{\mi}{\mathrm{i}}
\newcommand{\dif}{\mathrm{d}}
\newcommand{\Dif}{\mathrm{D}}

\newcommand{\sign}{\operatorname{sign}}
\renewcommand{\Im}{\operatorname{Im}}

\title{Collisions and Regularization for the 3-Vortex Problem}
\date{This version 9-XII-2004}
\author{Antonio Hern\'andez-Gardu\~no\thanks{%
	I.I.M.A.S., U.N.A.M.  M\'exico D.F.,
	C.P. 01000, M\'exico.
} \thanks{%
	Email:  ahernandez@leibniz. iimas.unam.mx\hspace{0.1em}.
}
\and 
Ernesto A. Lacomba\footnotemark[1] \thanks{%
	On leave of absence from Dpto. de Matem\'aticas, U.A.M.-Iztapalapa.  M\'exico D.F., C.P. 09340, M\'exico.  Email: lace@xanum.uam.mx\hspace{0.1em}.
}}

\begin{document}

\maketitle

\begin{abstract}
	We study the dynamics of 3 point-vortices on the plane for a fluid governed by Euler's equations, concentrating on the case when the moment of inertia is zero.  We prove that the only motions that lead to total collisions are self-similar and that there are no binary collisions.  Also, we give a regularization of the reduced system around collinear configurations (excluding binary collisions) which smoothes out the dynamics.
\end{abstract}

\section{Introduction}

The dynamical equation for the vorticity field $\omega$ on an ideal incompressible fluid with no viscosity is derived from Euler's equation and is given by $ \Dif \omega / \Dif t = \omega \cdot \nabla u $.  Here $u$ is the velocity field, $\omega=\nabla \times u$ is the vorticity field and $ \Dif / \Dif t := \partial / \partial t + u \cdot \nabla $ is the material derivative.  The context of this paper is the situation when the vorticity field $\omega$ is highly localized around a finite number of points, in effect being described by a delta function.  This is known as the $N$-point-vortex (or $N$-vortex, for brevity) problem. Each point-vortex is completely characterized by its position and its vortex-intensity. It is analogous to the $N$-point-masses considered in celestial mechanics, except that, in contrast to the masses, the vortex-intensities can be positive or negative.  We will concentrate on the case when the motion takes place on the plane.  This corresponds to the physical situation of $N$ parallel vortex filaments on three-dimensional space.

The equations of motion for the $N$-vortex problem on the plane, given by \eqref{vortex eqns of motion}, are derived in most standard texts on hydrodynamics, e.g. \cite{Batchelor2000}.  It is a Hamiltonian system.  Its solution for $ N = 2 $ is a routine exercise.  For $ N \ge 4 $ it is a non-integrable system.  The case $ N = 3 $ is interesting because it is still integrable and it has been shown to be a fundamental component of the $N$-vortex problem for large $N$.    

The study of the $3$ point-vortex problem was pioneered by W. Gr\"obli in his 1877 dissertation\footnote{%
	Cf. \cite{ArefRottThomann1992} for a detailed account on Gr\"obli's work.
}, who established the integrability of the system and obtained analytical solutions for certain special cases.  Several decades later, J. L. Synge \cite{Synge1949} studied the dynamics of the reduced system obtained by ignoring rotations and dilations, giving a classification of the possible reduced dynamics based on the harmonic mean (the virial) of the vorticities.  For this he used a geometric method based on using the so called \emph{trilinear coordinates} and an argument based on the linearization around the equilateral configuration.  Later, E. A. Novikov \cite{Novikov1975} gave an account of the solutions to the problem with three equal vorticities and, with Yu. B. Sedov \cite{NovikovSedov1979}, studied self-similar collapse (defined below) and the dynamics near collapse when the harmonic mean of the vorticities is zero.  H. Aref \cite{Aref1979} gave a further classification of the possible motions according to the harmonic, geometric and arithmetic mean of the vorticities, bringing together the methods in \cite{NovikovSedov1979} and \cite{Synge1949}.  In another significant development, D. M. F. Chapman \cite{Chapman1978} studied the relation between conserved quantities and symmetries (translational, rotational, temporal and scale transformations).  Most of the results in these references are nicely summarized and discussed in \cite[\S 2.2]{Newton2001}, where some additional material can be found.

This paper deals with the $3$-vortex problem on the plane and has two main purposes.  The first one is to give a negative answer to the question of whether there exist non-self-similar collisions.  Surprisingly, this question has not been fully addressed in the literature.  (Aref's \cite{Aref1979} description of the dynamics for zero moment of inertia does imply the impossibility of non-self-similar trajectories leading to collision, but his presentation is based on illustrative examples and not a definite proof.)  
As a by-product of our answer to this question we offer a simple description of the dynamics for the case when the moment of inertia is zero (which contains the self-similar collisions when the virial is null).  This description necessarily overlaps with some results in the references mentioned above, but the method that we use, based on a projective blow-up, is different.  The second objective of this paper is to describe the construction of a coordinate system that regularizes the vector field of this system near collinear configurations.  This is another sort of blow-up which duplicates the phase space of a vector field around singular points where a square root becomes zero.  As part of this discussion we also show that for the three vortex problem the binary collisions are excluded from the dynamics.


The system of $N$ point-vortices on the plane can be described as follows.  Let $ z _\alpha = x _\alpha + \mi \, y _\alpha \in \mathbb{C} $ be the coordinates of the $ \alpha $-th point vortex with vorticity $ \Gamma _\alpha $, $ \alpha = 1, \ldots , N $.  Thus the configuration space is $ \mathbb{C} ^n \cong \mathbb{R} ^{ 2n} $.  The equations of motion are given by\footnote{%
	Cf. \cite[\S 1.1.4]{Newton2001}
}
\begin{equation}\label{vortex eqns of motion}
	\dot{ z} _\alpha = \frac{ \mi }{ 2 \pi} \sum _{ \beta \neq \alpha}^N \Gamma _\beta \frac{ z _\alpha - z _\beta }{ |z _\alpha - z _\beta | ^2 }
\end{equation}
This system is Hamiltonian, which can be seen as follows.  Give $ \mathbb{C} ^n $ the symplectic form $ \Omega $ defined by
\[
  \Omega (z, w) = - \operatorname{ Im} \sum _{ \alpha = 1} ^n \Gamma _\alpha z _\alpha \bar{ w} _\alpha \;, 
\] 
where the $ \Gamma _\alpha \in \mathbb{R} $, $ \Gamma _\alpha \neq 0 $.  Then Hamilton's equation $ \mathbf{ i} _{ X _H } \Omega = \mathbf{ d} H $ takes the form 
\begin{equation}\label{general Hams eqns}
	\Gamma _\alpha \dot{ z} _\alpha = - 2 \mi \frac{ \partial H }{ \partial \bar{ z} _\alpha } \;, \quad \alpha = 1, \ldots N \;.  
\end{equation}
Let 
\[
  H = - \frac{ 1 }{ 2 \pi } \sum _{ \alpha < \beta} \Gamma _\alpha \Gamma _\beta \ln l _{ \alpha \beta} \;, 
\] 
where $ l _{ \alpha \beta} := | z _\alpha - z _\beta | $.  It is easy to see that equations \eqref{general Hams eqns} yield \eqref{vortex eqns of motion} and hence the system has a Hamiltonian structure.  In particular $H$ (the total energy) is a conserved quantity.  From \eqref{vortex eqns of motion} it follows that also
\[
  Z = \sum _\alpha \Gamma _\alpha z _\alpha 
\] 
and
\[
  I = \sum _\alpha \Gamma \rho _\alpha ^2 \;, \quad \rho _\alpha = | z _\alpha |
\] 
are conserved quantities.

Let $ \Gamma = \sum _\alpha \Gamma _\alpha $.  If $ \Gamma \neq 0 $ we call $ Z/\Gamma $ the barycenter of the system.  A calculation shows that
\[
  M := \sum _{ \alpha < \beta} \Gamma _\alpha \Gamma _\beta \, l _{ \alpha \beta} ^2 = \Gamma I - |Z| ^2 \;. 
\] 
Finally, from \eqref{vortex eqns of motion} it also follows that the angular momentum or \textbf{virial}
\[
  V = \sum _\alpha \Gamma _\alpha \, z _\alpha \times \dot{z}_\alpha =\sum _{\alpha<\beta} \Gamma _\alpha \Gamma _\beta
\] 
is an invariant of the system, i.e., a constant depending only on the vorticities.
Without loss of generality we will assume that $ Z \equiv 0 $ when $\Gamma\neq 0$, and hence $ M = \Gamma I $.  

Notice that all initial configurations that end in collision must satisfy $ M = I = 0 $ (even when $\Gamma = 0$).  This can be viewed as an extension to the three vortex problem of Sundman's theorem from celestial mechanics, which states that total collision implies that the angular momentum must be zero, see \cite{Pollard1976}.

We say that the dynamical evolution of $N$ vortices on the plane is a \textbf{self-similar} trajectory if it can be expressed as $ z _i (t) = \zeta (t) z _i (0) $, $ i = 1, \ldots N $, for some $ \zeta : \mathbb{R} \longrightarrow \mathbb{C} $ (hence $ Z = 0 $).  As we will see below, a self-similar trajectory that is not a \emph{relative equilibrium} (defined at the end 
of section \ref{sec:collision trajectories}) leads to total collision or ejection.

\section{Reduced coordinates and trajectories leading to total collision}
\label{sec:collision trajectories}

A given configuration of $N$ point-vortices can be recovered if one knows all the distances $ l _{ \alpha \beta} $ between the vortices and the overall orientation of the system.  The reduced system obtained after ignoring the orientation is called the \textbf{shape configuration}.  This will be enough for studying candidate collision trajectories.

From now on we will concentrate on the case $ N = 3 $.  
The way we prove the main result of this section, stating that any collision of planar 3-point-vortices is always self-similar, is by writing the conditions of collapse in terms of coordinates introduced by Novikov \cite{NovikovSedov1979}, which are essentially mutual distances, without requiring that the virial (or angular momentum) vanishes.  We prove that the cases of positive or negative virial leads to no solutions where all the 3 vortices approach the origin.  For this we apply a projective blow-up on a plane in the space of parameters.  The only case left is that of null virial, which is known to contain the solutions of self-similar collapse.

Let $ C _3 $ denote the cyclic permutations of $ (1,2,3) $ and let us introduce the notation $ b _i := l _{ jk} ^2 $, $ (i,j,k) \in C _3 $.

The shape configuration of the system is a point $ (b _1, b _2, b _3) $ in the first octant of $ \mathbb{R} ^3 $ such that the triangle inequalities
\begin{equation}\label{triangle ineq}
  \sqrt{b _i} + \sqrt{b _j} \ge \sqrt{b _k} \;, \quad (i,j,k) \in C _3 
\end{equation} 
are satisfied.  These three inequalities are equivalent to the single inequality,
\begin{equation}\label{cone eqn}
  b _1 ^2 + b _2 ^2 + b _3 ^2 \le  2(b _1 b _2 + b _1 b _3 + b _2 b _3) \;. 
\end{equation}
Geometrically, the set of points $ (b _1, b _2, b _3) \in \mathbb{R} ^3 $ satisfying $ b _i \ge 0 $ and \eqref{cone eqn} is a solid cone which we call the \textbf{admissible cone}, with symmetry axis the line $ b _1 = b _2 = b _3 $; this is clear from the symmetry and homogeneity of \eqref{cone eqn}.

Now we study the trajectories determined by the constraint that $H$ and $M$ must be conserved quantities.  
As it was mentioned in the introduction, all candidate trajectories for total collision must satisfy $ M = 0 $.  Another condition for total collision is that not all the $ \Gamma _i $ have the same sign; for if that were the case then $H$ would tend to $ \infty $ as $ l _{ \alpha \beta} \rightarrow 0 $.  For concreteness let us assume that $ \Gamma _1 \Gamma _2 > 0 $, $ \Gamma _1 \Gamma _3 < 0 $ and thus $ \Gamma _2 \Gamma _3 < 0 $.

Let $ \mathbf{ b} (t) = (b _1 (t), b _2 (t), b _3(t)) $ be a dynamical trajectory of the system satisfying \eqref{vortex eqns of motion}.  Since $H$ and $M$ are conserved, we have that, for all $t$,
\[\begin{split}
	H &= - \frac{ 1 }{ 4 \pi} \left( \Gamma _1 \Gamma _2 \ln b _3 + \Gamma _2 \Gamma _3 \ln b _1 + \Gamma _3 \Gamma _1 \ln b _2 \right) \;, \\
	0 &= \Gamma _1 \Gamma _2 b _3 + \Gamma _2 \Gamma _3 b _1 + \Gamma _3 \Gamma _1 b _2 \;. 
\end{split}\]
Exponentiating the first equation we obtain the equivalent system
\begin{subequations}\label{conservation constraint}
\begin{gather}
	\label{conservation constraint a}
	b _3 = h\, b _1 ^m b _2 ^n \;, \\[1.5ex]
	\label{conservation constraint b}
	m b _1 + n b _2 - b _3 = 0  
\end{gather}
\end{subequations}
where $ m = - \Gamma _3 / \Gamma _1 $, $ n = - \Gamma _3 / \Gamma _2 $, $ h = e ^{ -4 \pi H / (\Gamma _1 \Gamma _2) }  > 0 $.  Equation \eqref{conservation constraint a} expresses conservation of energy and equation \eqref{conservation constraint b} the condition $ M = 0 $.  Notice that $ m, n $ and $h$ are strictly positive.

The dynamics of the system is contained in the region $\mathcal{T} = \mathcal{C} \cap \{ M = 0 \} $, where $\mathcal{C}$ is the admissible cone determined by \eqref{cone eqn} and $ \{ M = 0 \} $ denotes the plane determined by \eqref{conservation constraint b}.  If $\mathcal{T} \neq \varnothing $ then $\mathcal{T}$ is a planar region contained in the first octant of $ \mathbb{R} ^3 $ bounded by two rays $ r _1, r _2 $ through the origin (possibly $ r _1 = r _2 $).  The projection of these rays on the $x$-$y$ plane determine two rays with slopes $ p _1 $, $ p _2 $ such that $ 0 \leq p _1 \leq p _2 \leq \infty $.  Its values are obtained by eliminating $ b _3 $ from the equality case in \eqref{cone eqn} and equation \eqref{conservation constraint b}, which gives the quadratic equation in $ p = b _2/b _1 $,
\begin{equation}\label{boundary line eqn}
	(n-1)^2 p ^2 + 2 (m n - m - n - 1) p + (m-1) ^2 = 0 \;, 
\end{equation}
whose roots are $ p _{ 1,2} = (1 \pm \sqrt{ \beta} ) ^2 / (n-1) ^2 $, where
\[
  \beta := m + n - m n \;. 
\]
For $ \beta < 0 $, $ \beta = 0 $ and $ \beta > 0 $, with $ \beta \neq 1 $, equation \eqref{boundary line eqn} has no real roots, exactly one positive root and exactly two positive roots, respectively.  
Notice that $ \beta = 1 $ iff $ m = 1 $ or $ n = 1 $, which corresponds to \eqref{boundary line eqn} having a root $ p = 0 $ or $ p = \infty $, respectively.  
Thus, $\mathcal{T}$ is not empty iff $ \beta \ge 0 $, it degenerates to a line when $ \beta = 0 $ and has a boundary line on the cartesian plane $x$-$z$ or $y$-$z$ iff $ m = 1 $ or $ n = 1 $, respectively.

Let 
\begin{equation}\label{m n virial}
	\gamma = 1 - m - n = \Gamma _3 \left( \frac{ 1 }{ \Gamma _1} + \frac{ 1 }{ \Gamma _2} + \frac{ 1 }{ \Gamma _3} \right) \;.
\end{equation}
We now consider the two possible cases:

\paragraph{Case $ \gamma = 0 $.}  In this case $ V = 0 $, which corresponds to the well known self-similar collisions described in \cite{Aref1979}.  Notice that in this case $ 0 < \beta < 1 $, and that $\mathcal{T}$ contains the symmetry axis of $\mathcal{C}$.  From the homogeneity of (\ref {conservation constraint}a) it is clear that for each ray $r$ through the origin contained in $\mathcal{T}$ there is an energy value $h$ such that $r$ is a set of solutions of \eqref{conservation constraint}.  Therefore, the region $\mathcal{T}$ is foliated by 
rays through the origin which are the dynamical orbits of the system, except for the boundary rays $r_1$, $r_2$ and the bisector of $\mathcal{T}$, which consist of equilibrium points, as we show in section \ref{sec:relative equilibria}.

\paragraph{Case $ \gamma \neq 0 $.}  Letting $ x = b _1 $, $ y = b _2 $, $ z = b _3 $ we obtain from equations \eqref{conservation constraint} the equivalent conditions
\begin{equation}\label{conservation constraint 2}
\begin{split}
	(m + n p) x ^{ 1 - m - n} - h p ^n = 0 \;, \\[1ex]
	y = p x \;, \quad z = m x + n y \;. 
\end{split}
\end{equation}
Therefore,
\begin{equation}\label{parametrized solution}
  x = \frac{ h ^{ 1/\gamma} p ^{ n/\gamma} }{ (m + n p) ^{ 1/\gamma }} \;, \quad
  y = \frac{ h ^{ 1/\gamma} p ^{ (1-m)/\gamma} }{ (m + n p) ^{ 1/\gamma }} \;, \quad
  z = \frac{ h ^{ 1/\gamma} p ^{ n/\gamma} }{ (m + n p) ^{ (m + n)/\gamma}} \;.
\end{equation}
Equations \eqref{parametrized solution} describe a parametrized curve $ c(p) = (x(p), y(p), z(p) $ defined and different from zero for all $ p \in (0, \infty) $.  If $ \gamma > 0 $ then $ (x, y, z) \rightarrow 0 $ as $ p \rightarrow 0 $ or $ p \rightarrow \infty $ (hence the curve is bounded).  If $ \gamma < 0 $ we see that, if $ p \rightarrow 0 $, respectively $ p \rightarrow \infty $, then $z$ behaves asymptotically as $ p ^{ n/\gamma} $, respectively $ p ^{ -m/\gamma} $.  In either case $ z \rightarrow \infty $ as $ p \rightarrow 0 $ or $ p \rightarrow \infty $ (hence the curve is unbounded).

The dynamical orbit is contained in the portion of $c$ lying on the planar region $\mathcal{T}$.  As we will show in section \ref{sec:relative equilibria}, the only equilibrium points occur when $ \beta = 0 $ and in this case $\mathcal{T}$ degenerates to a line of equilibrium points.  
If $ \beta \neq 0 $ and $ \beta \neq 1 $ then, from the discussion following equation \eqref{boundary line eqn}, the dynamical orbit consists of the curve given by \eqref{parametrized solution} with $ 0 < p _1 \leq p \leq p _2 < \infty $.  
This is a simple curve away from the origin joining two points, one on each boundary ray of $\mathcal{T}$, with the property that each ray through the origin in $\mathcal{T}$ intersects the curve exactly once.  
If $ \beta = 1 $ then, since in this case $ m = 1 $ or $ n = 1 $, $ \gamma < 0 $ and the curve does not tend to the origin as $ p \rightarrow 0 $ or $ p \rightarrow \infty $.

We have thus proved:

\begin{theorem}
	For a system of three point-vortices on the plane all total collisions are self-similar. They occur only when vorticities satisfy $V=0$ and they lie on the level $M=0$. 
\end{theorem}

\paragraph{Remark.}  The parametrization of the dynamical orbit by the variable $ p $ can be interpreted as a projective blow-up in the $x$-$y$ plane.

\paragraph{Remark.}  The equilibrium configurations $ \dot{b} _1 = \dot{b} _2 = \dot{b} _3 = 0 $ correspond to rigid motions of the vortices (all mutual distances are preserved).  It is usual to call these motions \textbf{relative equilibria}.

\section{Regularization of collinear configurations}
\label{sec:regularization}

In this section we introduce new coordinates that regularize the vector field at the boundary of the cone $\cal{ C}$ given by \eqref{cone eqn}.  Also, we show that it is not possible that two vortices collide while the third one remains away from the collision.  (Hence, the only kind of collision for three planar vortices is self-similar collapse.)

First we obtain\footnote{%
	See \cite{Synge1949} for a geometric derivation of these evolution equations.
} the evolution equations for the $ b _i $'s.  Let $ w _i = z _j - z _k $, hence $ b _i = |w _i | ^2 $.  Here and in what follows $ (\alpha, \beta, \gamma ) $ is a cyclic permutation of $ (1, 2, 3) $.  From the equations of motion \eqref{vortex eqns of motion} we get
\begin{equation}\label{walphadot}
	\dot{ w} _i = \frac{ \mi }{ 2 \pi } \left( \frac{ - \Gamma _i }{ \bar{ w} _k } + \frac{ \Gamma _k }{ \bar{ w} _i } - \frac{ \Gamma _i }{ \bar{ w} _j } + \frac{ \Gamma _j }{ \bar{ w} _i } \right)
\end{equation}
Thus,
\[\begin{split}
  \dot{ b} _i &= \frac{\dif}{\dif t} \left( w _i \bar{ w} _i \right) = \dot{ w} _i \bar{ w} _i + w _i \dot{ \bar{ w}} _i \\
  &= - \frac{ \Gamma _i }{ \pi} \Im \left( \frac{ w _i }{ w _j } + \frac{ w _i }{ w _k } \right) \;. 
\end{split}\] 
Noticing that $ w _i = -(w _j + w _k ) $,
\[\begin{split}
	\dot{ b} _i &= \frac{ \Gamma _i }{ \pi} \Im \left( \frac{ (w _j + w _k)^2 }{  w _j w _k } \right) = \frac{ \Gamma _i }{ \pi} \Im \left( \frac{ w _j }{ w _k } + \frac{ w _k }{ w _j } \right) \\
	&= \frac{ \Gamma _i }{ \pi} \Im \left( \bar{ w} _j w _k \right) \left( \frac1{| w _j | ^2} - \frac1{| w _k | ^2} \right) \;. 
\end{split}\] 
In the last equality we have used the identity
\[
  \Im \left( \frac{ z _1 }{ z _2 } + \frac{ z _2 }{ z _1 } \right) \equiv \Im \left( \bar{ z} _1 z _2 \right) \left( \frac1{ | z _1 | ^2} - \frac1{ | z _2 | ^2 } \right)\;, 
\] 
for all $ z _1, z _2 \in \mathbb{C} \setminus \{0\} $, which is easily proved by expressing the $ z _i $ in polar coordinates.  Let $A$ denote the oriented area of the triangle formed by the point vortices located at $ z _1, z _2, z _3 \in \mathbb{C} $.  It is easy to see that $ 2A = \Im \bar{ w} _j w _k $.  Hence,
\begin{equation}\label{bi_evolution eqns}
	\dot{ b} _i = \frac{ \Gamma _i A }{ 2 \pi} \left( \frac1{ b _j } - \frac1{ b _k} \right) \;, \quad \text{$(i, j, k)$ cyclic.}
\end{equation}

This represents in fact two vector fields in the interior of $\mathcal{C}$, since $A$ is defined up to a sign.  It becomes zero at the boundary of $\mathcal{C}$, but in a singular way since the expression of $A$ in terms of the $ b _i $'s involves the square root of a function equal to zero at $ \partial \mathcal{C} $ (cf. equation \eqref{area} below).  Thus, even though the original equations of motion \eqref{vortex eqns of motion} are regular around generic collinear configurations, the equations of motion \eqref{bi_evolution eqns} on the reduced space of shape configurations are not.

The vector field defined by \eqref{bi_evolution eqns} becomes infinite when $ b _i \rightarrow 0 $ while $ b _j, b _k $ remain away from zero, for some $ (i,j, k) \in C _3 $.  But we now show that such a \textbf{binary collision} can not be reached.  Indeed, assume that $ b _i \rightarrow 0 $ as $ t \rightarrow t _0 $, $ t _0 \leq \infty $.  From the triangle inequalities \eqref{triangle ineq} we have that $ \sqrt{ b _i} \geq | \sqrt{ b _j} - \sqrt{ b _k} | $ and hence $ | b _j - b _k | \rightarrow 0 $.  Expressing the conservation of energy as
\begin{equation}\label{conservation h}
	h = b _1 ^{ 1 / \Gamma _1} b _2 ^{ 1/\Gamma _2} b _3 ^{ 1/\Gamma _3} 
\end{equation}
we see that $ b _j, b _k $ can not tend to $c$ with $ 0 < c < \infty $.  Therefore either $ b _j, b _k \rightarrow 0 $ or $ b _j, b _k \rightarrow \infty $, as $ t \rightarrow t _0 $.  The former corresponds to total collision, the kind considered in the previous section, and by definition is not a binary collision; so let us assume the latter.  Conservation of $M$ can be written as
\[
  M = \Gamma _i (\Gamma _j + \Gamma _k ) b _k + \Gamma _j \Gamma _k b _i + \Gamma _k \Gamma _i (b _j - b _k ) \;. 
\] 
Since the last two terms in the right-hand-side tend to zero while $ b _k \rightarrow \infty $ we conclude that $ M = 0 $ and $ \Gamma _j + \Gamma _k = 0 $.  Therefore \eqref{conservation h} can be rewritten as
\[
  b _i ^{ 1/\Gamma _i} = h \left( 1 + \frac{ b _j - b _k }{ b _k } \right) ^{ 1/\Gamma _k} \;. 
\] 
The left-hand-side tends to $0$ or $ \infty $, while the right-hand-side tends to $ h > 0 $, which is a contradiction.  We have thus proved:

\begin{theorem}\label{theo:no binary collisions}
	The dynamical trajectory of any three-point-vortex configuration does not evolve into a binary collision.
\end{theorem}

Now we introduce coordinates $ ( \alpha, \lambda, \theta ) $ which, as we will show, regularize the vector field at the boundary of the admissible cone \eqref{cone eqn}.  The price that we will pay is that in the new coordinates the vector field at the equilateral configurations (which, as we show in section \ref{sec:relative equilibria}, are equilibria) will become undefined.  These coordinates are defined as follows.
Let 
\begin{equation}\label{lambda}
	\lambda = b _1 + b _2 + b _3 \;, \quad b _i = | z _j - z _k | ^2 \;. 
\end{equation}
Consider three points on the plane \eqref{lambda} for a fixed $\lambda$: $P _2 = (\lambda /2, \lambda /2, 0) $, $ P _1 = (\lambda/3, \lambda/3, \lambda/3) $ and $ \mathbf{b} = (b _1, b _2, b _3) $.  Thus $ P _1 $ is the center of the disc $ \mathcal{D} $ on the plane \eqref{lambda} bounded by the cone \eqref{cone eqn} and $ \stackrel{\longrightarrow}{P _1 P _2} $ is the radial segment of $\mathcal{D}$ perpendicular to the $ b _1 b _2 $-plane.  Let $\theta$ be the angle needed to take the ray $ \stackrel{\longrightarrow}{P _1 P _2} $ to $ \stackrel{\longrightarrow}{P _1 \mathbf{b}} $.
Finally, let $ \alpha = 12 A/( \sqrt3 \lambda) $.

From Heron's formula for the area of a triangle,
\begin{equation}\label{area}
\begin{split}
	4 A &= \varepsilon \sqrt{2(b _1 b _2 + b _1 b _3 + b _2 b _3) - b _1 ^2 - b _2 ^2 - b _3 ^2} \\
	&= \varepsilon \sqrt{\lambda ^2 - 2(b _1 ^2 + b _2 ^2 + b _3 ^2 )} \;, 
\end{split}
\end{equation}
where $ \varepsilon = \pm 1 $, depending on the orientation of the triangle formed by the point-vortices.  It is clear that $ |A| $ takes its maximum value when $ b _1 = b _2 = b _3 = \lambda / 3 $ so that $ |A| _{\text{max}} = \sqrt3 \, \lambda /12 $.  Let
\[
  \mathcal{M} := (-1,1) \times \mathbb{R} ^{+} \times S ^1 = \mathcal{M} _{-} \sqcup \mathcal{M} _0 \sqcup \mathcal{M} _+
\]
where
\[
  \mathcal{M} _{ -} := (-1,0) \times \mathbb{R} ^{+} \times S ^1 \;, \quad 
  \mathcal{M} _+ := (0,1) \times \mathbb{R} ^{+} \times S ^1
\]
and
\[
  \mathcal{M} _0 := \{0\} \times \mathbb{R} ^{+} \times S ^1
\] 
is the common boundary of $ \mathcal{M} _{-} $ and $ \mathcal{M} _{+} $.  It is easy to see that the correspondence between the variables $ (\alpha, \lambda, \theta) $ and $ (b _1, b _2, b _3) $ induces a diffeomorphism between $ \mathcal{M} _\pm $ and $\mathcal{C} \setminus \mathcal{E} $, where $C$ is the admissible cone defined by \eqref{cone eqn} and 
\[
  \mathcal{E} := \{ (b _1, b _2, b _3) \in \mathcal{C} \mid b _1 = b _2 = b _3 \} \qquad \text{(equilateral configurations)} \;,
\]
and a diffeomorphism between $ \mathcal{M} _0 $ and $ \partial \mathcal{C} \setminus \mathcal{E} $ (these are the collinear configurations minus total collapse).   In contrast, the boundary $ \partial \mathcal{M} $ corresponds, in a non-bijective manner, to the equilateral configurations $ \mathcal{E} $.  (Notice that the total collision belongs to $\mathcal{E}$.)

Let us obtain the expressions of the $ b _i $ as functions of $ \alpha, \lambda, \theta $.  From \eqref{area} and the definition of $\alpha$ we have that
\begin{equation}\label{b-hypothenuse}
	b _1 ^2 + b _2 ^2 + b _3 ^2 = \frac{ \lambda ^2 }{ 6 } (3 - \alpha ^2 ) \;. 
\end{equation}
Also,
\[
	\cos \theta = \frac{ \stackrel{\longrightarrow}{P _1 \mathbf{b}} \cdot \stackrel{\longrightarrow}{P _1 P _2} }{ | \stackrel{\longrightarrow}{P _1 \mathbf{b}} | \, | \stackrel{\longrightarrow}{P _1 P _2} | } = \frac{ b _1 + b _2 - 2 b _3 }{ \sqrt6 \sqrt{b _1 ^2 + b _2 ^2 + b _3 ^2 - \lambda ^2 / 3} } \;. 
\] 
Using that $ \sin ^2 \theta = 1 - cos ^2 \theta $ and that $ \sign (\sin \theta) = \sign (b _2 - b _1) $ we get that 
\[
  \sin \theta = \frac{ b _2 - b _1 }{ \sqrt2 \sqrt{ b _1 + b _2 + b _3 - \lambda ^2/3} } \;. 
\] 
Using \eqref{b-hypothenuse} these expressions can be simplified:
\begin{equation}\label{trigonometric-theta}
  \cos \theta = \frac{ \lambda - 3 b _3 }{ \lambda \sqrt{1 - \alpha ^2} } \;, \quad \sin \theta = \frac{ \sqrt3 (b _2 - b _1) }{ \lambda \sqrt{ 1 - \alpha ^2} } \;. 
\end{equation}
From \eqref{lambda} and \eqref{trigonometric-theta} we obtain a linear system for the $ b _i $'s whose solution is
\begin{equation}\label{b-linear system}
\begin{split}
	b _1 &= \frac \lambda 6   \left( 2 + \sqrt{1-\alpha ^2} (\cos \theta - \sqrt3 \sin \theta ) \right) \;, \\
	b _2 &= \frac \lambda 6   \left( 2 + \sqrt{1-\alpha ^2} (\cos \theta + \sqrt3 \sin \theta ) \right) \;, \\
	b _3 &= \frac \lambda 3   \left( 1 - \sqrt{1- \alpha ^2} \cos \theta \right) \;, 
\end{split}
\end{equation}
so that the $ b _i $'s are differentiable functions of $ (\alpha, \lambda, \theta) \in \mathcal{M} $.

From \eqref{bi_evolution eqns}, \eqref{lambda} and the definition of $\alpha$ we see that $ \dot{ \lambda} = \alpha f _\lambda $, where $ f _\lambda $ is a rational function of the $ b _i $'s, with parameters $ \Gamma _i $.  From \eqref{bi_evolution eqns} and \eqref{area} we get that
\begin{multline}\label{area dot}
  \dot{A} = \frac1{32 \pi} \Bigg[ \Gamma _1 \left( \frac1{b _2} - \frac1{b _3} \right) (-b _1 + b _2 + b _3) + \Gamma _2 \left( \frac1{b _3} - \frac1{b _1} \right) (b _1 - b _2 + b _3) \\
  + \Gamma _3 \left( \frac1{b _1} - \frac1{b _2} \right) (b _1 + b _2 - b _3) \Bigg] \;. 
\end{multline}
Hence, $ \dot{\alpha} = (12 f _A / \sqrt3 - \alpha ^2 f _\lambda ) / \lambda $, where $ f _A $ is another rational function of $ b _1, b _2, b _3$ with parameters $ \Gamma _i $.  Both $ f _\lambda $ and $ f _A $ are differentiable except when $ b _i = 0 $ for some $i$.  Hence the same is true for $ \dot{ \lambda} $, $ \dot{ \alpha} $ and, because of \eqref{trigonometric-theta} and \eqref{b-linear system}, it is also true for $ \dot{ \theta} = \cos \theta (\sin \theta) ^{\displaystyle\cdot} - \sin \theta (\cos \theta) ^{\displaystyle\cdot} $.

Therefore the variables $ ( \alpha, \lambda, \theta ) $ regularize the vector field \eqref{bi_evolution eqns} on $ \tilde{ \mathcal{M}} := \mathcal{M} \setminus \mathcal{B} $, where
\[
  \mathcal{B} = \{ \left( 0, \lambda, 2n \pi/3 \right) \in \mathcal{M} \mid n = 0, 1, 2 \} 
\] 
is the set of binary collisions, i.e. $ b _i = 0 $ for some $i$, which is a codimension $2$ closed subset of $\mathcal{M}$.  In this set the vector field is singular and, as stated in theorem \ref{theo:no binary collisions}, it can not be reached even asymptotically.

The vector field on $ \tilde{ \mathcal{M}} $ can be easily computed, as we now show.  From \eqref{b-linear system} one computes the total derivative $ \Dif (b _1, b _2, b _3)/ \Dif (\alpha, \lambda, \theta) $, whose determinant is equal to $ \sqrt3 \, \alpha \lambda ^2 / 18 $, so that it is invertible on $\mathcal{M}$ iff $ \alpha \neq 0 $.  Its inverse is computed to be
\[
  \frac{ \Dif ( \alpha, \lambda, \theta ) }{ \Dif ( b _1, b _2, b _3) } = \frac{ \sqrt{1 - \alpha ^2} }{ \alpha \lambda } \mathbf{ B} \;, 
\] 
where
\newcommand{\scs}{\scriptstyle}
\[\label{matriz}
  \mathbf{ B} := 
  \begin{pmatrix}
	\scs \sqrt{1 - \alpha ^2} + \sqrt3 \sin \theta - \cos \theta & \scs \sqrt{1 - \alpha ^2} - \sqrt3 \sin \theta - \cos \theta & \scs \sqrt{1-\alpha ^2} + 2 \cos \theta \\
	\scs \alpha \lambda / \sqrt{1-\alpha ^2} & \scs \alpha \lambda / \sqrt{1-\alpha ^2} & \scs \alpha \lambda / \sqrt{1-\alpha ^2} \\
	\frac{ -\alpha (\sqrt3 \sin \theta + 3 \cos \theta) }{ \sqrt3 (1- \alpha ^2) } & \frac{ -\alpha (\sqrt3 \sin \theta - 3 \cos \theta) }{ \sqrt3 (1-\alpha ^2) } & \frac{ 2 \alpha \sin \theta }{ 1-\alpha ^2}
  \end{pmatrix} \;. 
\] 
Thus, from \eqref{bi_evolution eqns} and the chain-rule,
\[
  \begin{pmatrix} \dot{ \alpha} \\ \dot{ \lambda } \\ \dot{ \theta} \end{pmatrix} 
  = \frac{ \sqrt3 \sqrt{ 1 - \alpha ^2} }{ 24 \pi} \mathbf{ B}
	\begin{pmatrix}
		\Gamma _1 (b _2 ^{-1} - b _3 ^{-1}) \\ 
		\Gamma _2 (b _3 ^{-1} - b _1 ^{-1}) \\ 
		\Gamma _3 (b _1 ^{-1} - b _2 ^{-1})
	\end{pmatrix} \;. 
\] 
where the $ b _i $ are given by \eqref{b-linear system}.  As expected, the vector field $ ( \dot{\alpha}, \dot{ \lambda}, \dot{ \theta} ) $ is regular on $ \tilde{ \mathcal{M}} $.  In particular, we have achieved the regularization of the vector field around the collinear configurations (excluding the non dynamical binary collisions).

\paragraph{Remark.}  It is clear that $ A = 0 $ iff $ \alpha = 0 $.  Noticing that from $\frac{\sqrt{3}}{12}\dot{\alpha}=\frac{\lambda \dot{A}-A\dot{\lambda}}{\lambda ^2}$
we see that when $ A = 0 $, we also have that $ \dot{ A} = 0 $ iff $ \dot{\alpha} = 0 $.

\section{Equilibrium configurations}
\label{sec:relative equilibria}

In this section we determine the equilibrium configurations of the system defined by equations \eqref{bi_evolution eqns}.  This completes the discussion in section \ref{sec:collision trajectories} where the description of the dynamical orbits requires distinguishing the equilibrium points.  As remarked in that section, these equilibria have the interpretation of being the relative equilibria of the 3-point vortex system.

Directly from \eqref{bi_evolution eqns} we see that a necessary condition for equilibrium is that either $ b _1 = b _2 = b _3 $ (equilateral configurations) or $ A = 0 $ (collinear configurations).  Since the vector field \eqref{bi_evolution eqns} is smooth around equilateral configurations, apart from total collision, then being at an equilateral configuration is also a sufficient condition for equilibrium.  On the other hand, $ A = 0 $ is not a sufficient condition for equilibrium because the expression of $A$ in terms of the $ b _i$'s involves a square root and hence the vector field is not smooth around collinear configurations.  In this case the equilibrium conditions at those configurations can be easily determined with the help of the regularized vector field $ ( \dot{ \alpha}, \dot{ \lambda}, \dot{ \theta} ) $ of section \ref{sec:regularization}:  from the expressions for $ (\dot{\alpha} , \dot{\lambda}, \dot{\theta} ) $, $ \mathbf{ B} $ and the remark at the end of section \ref{sec:regularization} 
we see that if $ A = 0 $ then a) $ \dot{\lambda} = \dot{\theta} = 0 $ and b) $ \dot{\alpha} = 0 $ iff $ \dot{A} = 0 $.  Therefore,

\begin{lemma}
	The equilibrium configurations of the evolution equations \eqref{bi_evolution eqns} are a) the equilateral configurations and b) the collinear configurations for which $ \dot{A} = 0 $.
\end{lemma}

We now determine the conditions for $ \dot{A} = 0 $ when $ A = 0 $ in the case when $ M = 0 $, which is the context of section \ref{sec:collision trajectories}.  First we observe that, from \eqref{conservation constraint b} and \eqref{area dot}, $ \dot{A} $ can be expressed in terms of $ p = b _2 / b _1 $ by
\begin{equation}\label{area dot bijk}
\newcommand{\margin}{\hspace{10ex}}
\left. \margin
\begin{array}{l}
	\displaystyle c\, \dot{A} = \frac 1m b _{ 123} + \frac 1n b _{ 231} - b _{ 312} \;, \\[1.6ex]
	\displaystyle b _{ ijk} := \left( \frac{ b _i }{ b _j} - \frac{ b _i }{ b _k} \right) \left( \frac{ b _j }{ b _i} + \frac{ b _k }{ b _i} -1 \right) \;, \\[1.6ex]
	\displaystyle \frac{ b _3 }{ b _1} = m + n p \;, \quad \frac{ b _3 }{ b _2} = m p ^{-1} + n \;, \quad \frac{ b _2 }{ b _1} = p \;, 
\end{array}
\margin \right\}
\end{equation}
where $ c = - 32 \pi / \Gamma _3 $ and $ (i,j,k) \in C _3 $.  The values of $p$ that correspond to $ A = 0 $ (the boundary lines of $\mathcal{T}$) are given by the roots of \eqref{boundary line eqn}.  A calculation shows that substituting these values of $p$ in \eqref{area dot bijk} and expressing $m$ and $n$ in terms of $ \beta = m + n - m n $ and $ \gamma = 1 - m - n $ yields 
\[
  c \dot{A} = \frac{ 2 \varepsilon \sqrt \beta \gamma }{ \beta + \gamma -1} \;. 
\] 
Therefore $ \dot{A} = 0 $ iff either $ \beta = 0 $ or $ \gamma = 0 $.  Noticing that $ \beta = - \frac{ \Gamma _3 }{ \Gamma _1 \Gamma _2 } \Gamma $ and recalling, from \eqref{m n virial}, that $\gamma$ is proportional to the virial, we conclude:

\begin{proposition}
	Assuming $ M = 0 $, a collinear configuration is an equilibrium if and only if either the total vorticity or the virial vanishes.
\end{proposition}

\paragraph{Remark.}  From the discussion following equation \eqref{boundary line eqn} we see that if the total vorticity vanishes then $\mathcal{T}$, the set of states satisfying $ M = 0 $, degenerates to a single line of equilibrium points.

\vspace{4ex}
We conclude by noticing that the self-similar trajectories of the case $ M = \gamma = 0 $ discussed in section \ref{sec:collision trajectories} correspond to collisions or ejections depending on the sign of $ \dot{ \lambda} $, where $ \lambda $ is one of the regularizing coordinates introduced in section \ref{sec:regularization}.  From the definition of $\lambda$ and equation \eqref{bi_evolution eqns} we compute that
\[
  \frac{ 2 \pi b _3 }{ \Gamma _3 A} \dot{\lambda} = \frac{ \left( (2-m)p + m + 1 \right) (p - 1) }{ p } \;,
\] 
where $ p = b _2 / b _1 $.
The right hand side is zero iff $ p = (m + 1) / (m - 2) $ or $ p = 1 $.  The former can not happen since it is incompatible with $ p > 0 $ and $ 0 < m < 1 $.  Also, for these ranges of $p$ and $m$, $ (2-m)p + m + 1 > 0 $.  Therefore, if $ A \neq 0 $, $ \dot{\lambda} = 0 $ iff $ p = 1 $; and if $ p \neq 1 $ then $ \sign ( \dot{\lambda}) = \sign [ \Gamma _3 \, A \, (p-1) ] = \sign [ \Gamma _3 \, \alpha \, \sin \theta ] $.  Thus, for a given orientation of the vorticities, the bisector of $\mathcal{T}$ (which corresponds to the equilateral configurations) separates the collapsing and expanding self-similar trajectories.  In terms of the regularizing coordinates, we see that the collapsing or expanding nature of the self-similar trajectories changes when $ \alpha \mapsto - \alpha $ or $ \theta \mapsto - \theta $.

\section*{Acknowledgments}

Both authors gratefully acknowledge support from DGAPA-UNAM under project PAPIIT IN101902 and from the Conacyt (Mexico) grant 32167-E. The second author wishes to thank the hospitality of the Department of Mathematical and Numerical Methods of IIMAS-UNAM during the period in which this paper was prepared.


\end{document}